\documentclass[12pt]{article}
\usepackage{times}
\usepackage{geometry}
\usepackage{graphicx}
\usepackage[utf8]{inputenc}
\usepackage{enumitem,amssymb}
\usepackage{ragged2e}
\usepackage[sort, numbers]{natbib}
\usepackage{pifont}
\newcommand{\cmark}{\ding{51}}%
\newcommand{\done}{\rlap{$\square$}{\raisebox{2pt}{\large\hspace{1pt}\cmark}}%
\hspace{-2.5pt}}

\begin{document}
\raggedright
\huge
Astro2020 Science White Paper \linebreak

Radio Pulsar Populations \linebreak
\normalsize

\noindent \textbf{Thematic Areas:} \hspace*{60pt} $\square$ Planetary Systems \hspace*{10pt} $\square$ Star and Planet Formation \hspace*{20pt}\linebreak
$\done$ Formation and Evolution of Compact Objects \hspace*{31pt} $\square$ Cosmology and Fundamental Physics \linebreak
  $\square$  Stars and Stellar Evolution \hspace*{1pt} $\square$ Resolved Stellar Populations and their Environments \hspace*{40pt} \linebreak
  $\square$    Galaxy Evolution   \hspace*{45pt} $\square$             Multi-Messenger Astronomy and Astrophysics \hspace*{65pt} \linebreak
  
\textbf{Principal Author:}

Name:    Duncan Lorimer
 \linebreak                        
Institution:  West Virginia University
 \linebreak
Email: duncan.lorimer@mail.wvu.edu
 \linebreak
Phone:  304-290-0417
 \linebreak
 
\textbf{Co-authors:} 
 \linebreak                        
Nihan Pol (West Virginia University)  \linebreak                
Kaustubh Rajwade (University of Manchester)\linebreak
Kshitij Aggarwal (West Virginia University)\linebreak
Devansh Agarwal (West Virginia University)\linebreak
Jay Strader (Michigan State University)\linebreak
Natalia Lewandowska (West Virginia University)\linebreak
David Kaplan (University of Wisconsin Milwaukee)
\linebreak
Tyler Cohen (New Mexico Tech)
\linebreak
Paul Demorest (National Radio Astronomy Observatory)
\linebreak
Emmanuel Fonseca (McGill University)
\linebreak
Shami Chatterjee (Cornell University)
\linebreak

\textbf{Abstract:}
\justifying
Our understanding of the neutron star population is informed to a great degree by large-scale surveys that have been carried out by radio facilities during the past fifty years. 
We summarize some of the recent breakthroughs in 
our understanding of the radio pulsar population, and look ahead to future yields from upcoming experiments planned for the next decade. By the end of the 2020s, we anticipate a much more complete census of the Galactic population and being able to probe the populations of radio-emitting neutron stars more effectively in external galaxies. Among the anticipated discoveries are pulsar--black hole binary systems that will provide further tests of strong-field gravity, as well as large numbers of millisecond pulsars that are crucial to enhancing the sensitivity of timing arrays for low-frequency gravitational waves.

\pagebreak

\section{Why Study Pulsar Populations?}

Radio pulsars, rapidly spinning highly magnetized neutron stars, have long been known as excellent tools to study fundamental physics and astronomy~\cite{bhlm92,lk05}. Of particular interest are ``millisecond pulsars'' (MSPs) which rotate with frequencies of several hundred Hz with atomic clock-like precision.  MSPs are frequently used as tools to investigate a variety of topics, including: the physics of matter at supranuclear densities~\cite{dpr+10,opr+10}, detection of low-frequency gravitational waves~\cite{jhv+07}, probes of general relativity~\cite{klb+05}, interstellar weather~\cite{rdb+06}, globular cluster astrophysics~\cite{cr05}, extrasolar planets~\cite{wf92,bbb+11} and planetary physics~\cite{chm+10}. Three of the many key astrophysical questions to be tackled in the upcoming decade through large-scale surveys and detailed modeling are outlined below.

\smallskip \noindent {\bf 1. How fast can a neutron star spin?}  On theoretical grounds, the maximum spin rate is governed by the density profile of neutron stars~\cite{lp04}. Softer profiles allow for the existence of more rapidly spinning neutron stars, and observations of any such objects would rule out the stiffer profiles.  X-ray observations \cite{cmm+03} suggest that the maximum spin frequency may be close to 700~Hz. Observationally, the fastest spinning neutron star is the 716~Hz pulsar in Terzan~5~\cite{hrs+06}. Do neutron stars exist with $>1$~kHz spin rates? Future surveys could definitively answer this question.

\smallskip \noindent {\bf 2. How many pulsars are there?} Studies of the spatial distribution of normal pulsars have shown that pulsars generally trace the radial distribution of massive stars in our Galaxy, but have a scale height above the plane of several hundred parsecs~\cite{lfl+06}. Although the vertical distribution of MSPs has been studied locally~\cite{lor95}, the radial density of MSPs is poorly understood. This topic is of great interest to: (i) our ongoing efforts \cite{mcl13} to optimize future surveys to find MSPs for gravitational wave detection and characterization by pulsar timing arrays; (ii) understanding the MSP contribution to the diffuse Galactic gamma-ray emission~\cite{fl10}.

\smallskip \noindent {\bf 3. How are pulsars formed?} Pulsars are superb probes of binary star evolution.  The existence of isolated MSPs and unusual binary/planetary systems implies that other progenitors are required to explain the MSP population we see beyond those predicted by the standard model. Of key importance is determining the birth rates of the various sub-samples of MSPs and comparing them to alternative formation routes (e.g.~the accretion-induced collapse of a white dwarf \cite{np89}). Many of the details of the assumptions (e.g.~mass transfer physics and neutron star birth velocities) are still not well understood. Improved knowledge of these areas would help studies of source classes for gravitational wave emission from compact binary inspirals~\cite{okf+05}.

Despite over 50 years of study, it is not well known how pulsars radiate. There is a wide variety of emission behavior from pulse to pulse through  nulling, mode-changing, giant pulses and subpulse drifting phenomena~\cite{bac70,bac70a,bac70b,bac70c,soglasnov_2004,staelin}.  Closely linked phenomena are the rotating radio transients \cite{mll+06} and the intermittent pulsars \cite{klo+06,crc+12,llm+13}. While the rotating radio transients emit only occasional individual pulses spaced anywhere between seconds to years, intermittent pulsars switch on and remain steady before switching off again over timescales of days to years. Over these longer timescales,
where it is possible to measure the spin period derivative precisely,
it is observed that the ``on'' and ``off'' states for the intermittent
pulsars are associated with two different spin-down rates! A similar
behavior was seen in 17 non-intermittent
pulsars \cite{lhk+10} which switch between different spin-down
states in an apparently quasi-periodic fashion, sometimes accompanied
by correlated pulse profile changes.  The mixture of spin-down states could even account for the timing noise
variations are seen in many pulsars and suggest that most, if not all,
normal pulsars have multiple spin-down states indicating global
changes in magnetospheric current densities.

\section{Modeling Pulsar Populations}

Inferring the properties of the underlying, as opposed to the observed, pulsar
population is a difficult problem. Not only does the detection process have to be accurately accounted for
to account for observational biases, the non-homogenous properties of the interstellar medium need to be
taken into account. Much of the progress has come from detailed Monte Carlo simulations of pulsar populations that create artificial pulsars that satisfy the criteria for detection, it is possible to generate and optimize models for the pulsar population which inform us about the underlying distribution functions and make predictions for future survey yields. Generally, the Monte Carlo simulations follow two basic strategies. In
the ``snapshot'' approach, no assumptions are made concerning the prior evolution of pulsars. Instead,  populations are generated according to various distribution functions (typically in Galactocentric radius, $R$, height with respect to the plane, $z$, spin period, $P$ and luminosity, $L$) which are optimized to find the best match to the sample. 

\begin{figure*}[h!]
\centering
\includegraphics[width=0.85\textwidth]{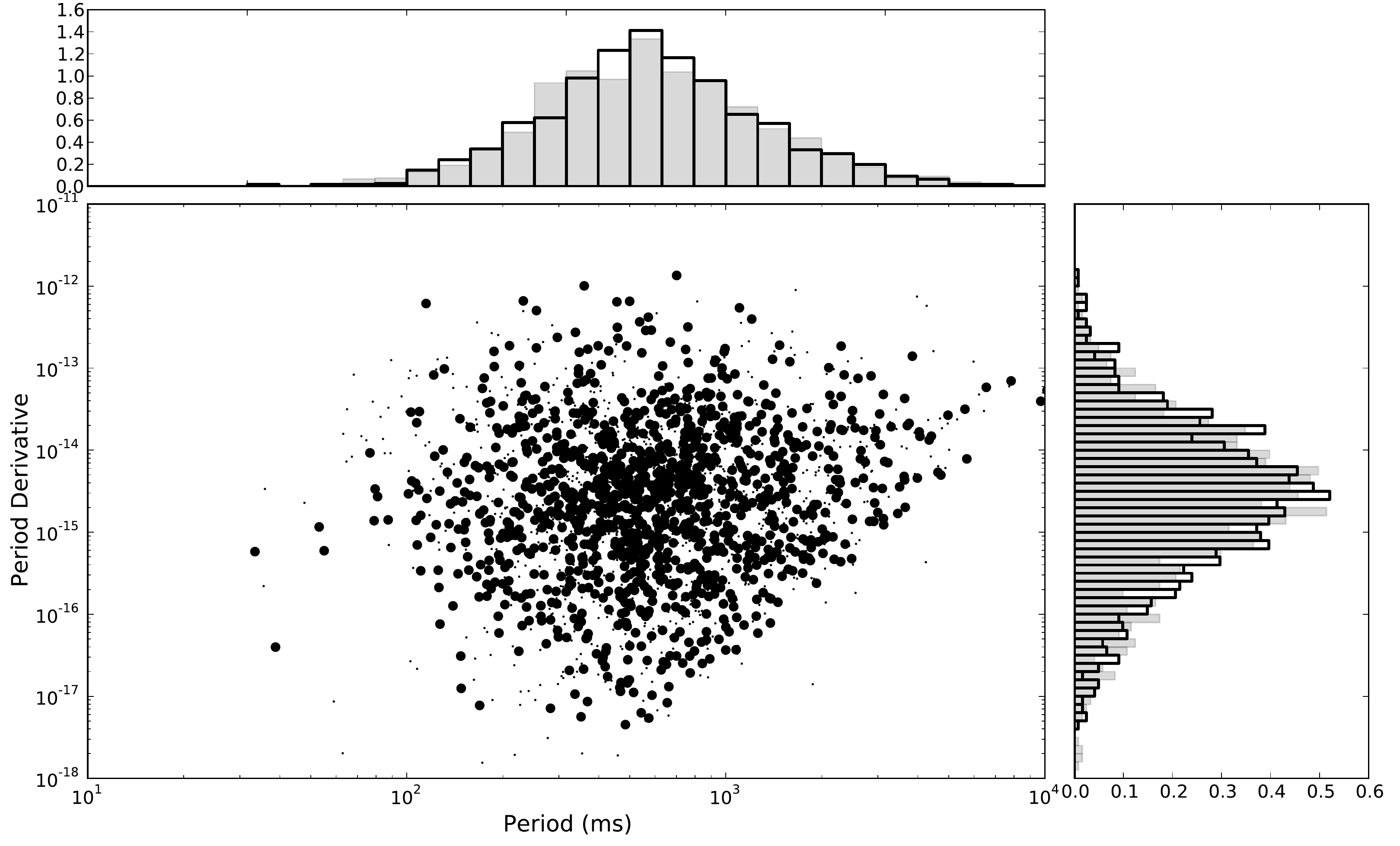}
\caption{$P$-$\dot{P}$ diagram showing Monte Carlo pulsars detected in models of surveys carried out with the Parkes
 telescope (bold points).
    $P$-$\dot{P}$ values that were taken from the pulsar catalog  (smaller points).
    Histograms of the $P$ and $\dot{P}$ distributions are also shown, with grey bars for catalog sources, and with solid lines for the simulated pulsars. Figure is taken from Bates et al.~\cite{blrs13}.
}

\end{figure*}

Alternatively, as shown in Fig.~1, one may carry out ``evolution'' approaches where the model pulsars are evolved forward in time from a set of initial distributions. Software to carry out both approaches has been developed by a number of groups and some of
this is freely available. For example, the {\tt PsrPopPy} software package \cite{blrs13} has modules to carry out both the snapshot and evolution approaches. Although many earlier studies did not attempt to rigorously optimize model parameters \cite{fk06}, in recent years these approaches have become more and more exhaustive and
employed Markov-Chain Monte Carlo methods to estimate
the density of pulsars in phase space and optimize these to find marginalized probability density functions for various model parameters \cite{2018arXiv180302397C,2018ApJ...863..199G} and to compute Bayes factors to compare different families of models.

\section{Pulsar Demography: Recent Advances}

Inspired by earlier studies \cite{tm77,lmt85}, the snapshot approach was applied to the normal pulsar population \cite{lfl+06}  to derive best-fitting probability
density functions in $R$, $L$, $z$ and $P$ for the present-day population of objects. One result of this work was that the radial
distribution of pulsars could not be decoupled from the radial
distribution of free electrons in the pulsar distribution.  For the
evolution approach on the normal population, a highly cited study is
the work of Faucher-Gigu\`ere \& Kaspi 
\cite{fk06} who
generated excellent fits to the pulsar $P$-$\dot{P}$ diagram (Fig.~1) using a
model in which the luminosity has a power-law dependence on $P$ and $\dot{P}$.
In addition,  the luminosity function of
the present-day pulsar population appears to be log-normal in form. 
The snapshot technique can be modified \cite{kkl03} such that it can be combined with the census of observed neutron star-white dwarf and double neutron star binary systems to calculate their respective merger detection rates for ground and space-based gravitational wave detectors \cite{ns_wd_merger_rate, B_merger_rate, Pol_merger_rate}.

Recent progress has been made in our understanding of the evolution of the magnetic
inclination angle, $\alpha$, for normal pulsars. Following earlier work \cite{2008MNRAS.387.1755W}, more detailed modeling
following the evolutionary framework outlined above by Johnston \& Karastergiou
\cite{2017MNRAS.467.3493J} demonstrated that an exponential decay of $\alpha$ on a $10^7$~yr time
scale can provide an excellent match to the $P$--$\dot{P}$ diagram for a pulsar
birth rate of one per century and an initial spin period of 20~ms. Historians
of pulsar statistics will be aware that the debate surrounding initial spins
of normal pulsars, their magnetic field and luminosity evolution, as well as
the role of $\alpha$ in the spin-down evolution model, has been a contentious issue. The new developments from this study, however, provide a framework for more accurately modeling
the evolution of the pulsar spin-down than has been previously possible. Theoretical progress in the past few years has arisen from detailed magnetohydrodynamical modeling
\cite{2006ApJ...648L..51S}. As a result, most of the pulsar evolution codes are now moving away from the simple magnetic dipole model \cite{go70} that has been used almost exclusively up until around 2010.

Our understanding of the Galactic population of MSPs and, more generally, recycled pulsars over the past decade has been significantly improved. This is thanks, in large part, to a great increase in sample size provided by a number of large-scale surveys of the Galactic disk at Parkes \cite{mlc+01}, Green Bank \cite{2018IAUS..337...13L}, Arecibo \cite{cfl+06},
Effelsberg \cite{2011AIPC.1357...52B} and the Low-Frequency Array \cite{2014A&A...570A..60C}. These surveys have greatly increased the sample size such that the number of Galactic MSPs now outnumbers their counterparts in globular clusters. A recent study by Gonthier et al.~\cite{2018ApJ...863..199G} combined radio surveys with results from the Fermi point source catalog \cite{2010ApJS..188..405A} to provide a comprehensive model of the radio and gamma-ray MSP population. 

Strong evidence for selection bias in MSPs comes from $\gamma$-ray observations with \emph{Fermi}. Among the most exciting discoveries of \emph{Fermi} has been that non-accreting pulsars are nearly ubiquitous emitters of GeV $\gamma$-rays \cite{2013ApJS..208...17A}. In addition to the physical insight possible by linking radio and high-energy observations of individual systems, radio timing observations of previously unassociated \emph{Fermi} sources have allowed the discovery of many new pulsars. In particular, there has been an enormous increase in the number of binary millisecond pulsars that ablate their hydrogen-rich companions. These are typically classified as either ``black widow'' or ``redback'' systems depending on whether the companion is less or more massive than $\sim 0.1 M_{\odot}$ \cite{2013IAUS..291..127R}. The pulsed radio emission from these systems is sometimes eclipsed due to ionized material associated with the companion, but this material apparently does not affect the detectability of the GeV emission. Thus, while existing pulsar catalogs are biased against systems with hydrogen-rich companions, follow-up of unassociated \emph{Fermi} sources has proven an effective way to discover and characterize these binaries \cite{2012arXiv1205.3089R}. 
These ``spider'' binaries are important to improve our understanding of the process by which pulsars are recycled, and to determine long-sought values such as the maximum mass of a neutron star. It is now clear that these binaries host more massive neutron stars than the canonical $1.4 M_{\odot}$, with median masses of $1.8 M_{\odot}$ \cite{2019ApJ...872...42S}, and several systems having minimum masses of $> 1.9 M_{\odot}$. The discovery and follow-up of more such binaries, which is best accomplished through optical, X-ray, $\gamma$-ray, and radio observations, offers one of the best paths to confirm neutron stars with masses $> 2 M_{\odot}$ to determine the maximum neutron star mass.

Two further outstanding questions are the population of pulsars in the inner Galaxy
and Galactic halo. In the central regions, there is intense interest in constraining this population both from the point of view of finding pulsars in orbit around
 SgrA* \cite{2016ApJ...818..121P} and from investigating whether the Galactic Center gamma-ray excess is due to an ensemble of millisecond pulsars \cite{2018MNRAS.480.4955F}. Due to the impacts of pulse dispersion and scattering, as well as the significant inverse square law distance penalty to the Galactic Center, searches have so far had limited sensitivity. A recent study taking these factors into account 
\cite{2017MNRAS.471..730R} suggests that the optimal frequency for future surveys of this region is 9--13~GHz. In the Galactic halo, a recent analysis shows that the fraction of the population bound therein depends critically on the magnetic angle evolution and birth velocity distribution \cite{2018MNRAS.479.3094R}.

The pulsar content in Galactic globular clusters has been studied extensively over the past thirty years. Recent studies have made use of constraints from the luminosity function inferred from the Galactic disk population to make predictions on the various types of pulsars. Highlights from this work include constraints on the highly anomalous population of young (non-recycled) pulsars in globular clusters \cite{blt+11} as well a census of the entire population \cite{tl13}. The key relationship between overall pulsar abundance appears to be the two-body encounter rate \cite{2010ApJ...714.1149H,tl13}. This leads to a number of predictions for cluster searches with new facilities now coming online.

Currently, the most distant pulsars known are in the Large and Small Magellanic Clouds \cite{mfl+06,2013MNRAS.433..138R}. The relatively small sample size at present has hampered attempts to model the pulsar content in these galaxies and has so far prohibited detailed comparisons with the Milky Way and globular cluster population. The next decade is likely to result in the discovery of a substantial number of extragalactic pulsars, which can be used to probe the Galactic halo and circumgalactic medium.

\section{Key Advances: Upcoming Surveys}

Many facilities are set to transform our understanding of the radio pulsar population over the next decade. Most notable are the Five Hundred Meter Aperture Radio Telescope (FAST), the Canadian H~\textsc{i} Intensity Mapping Experiment (CHIME), Hydrogen Intensity and Real-time Analysis eXperiment (HIRAX), MeerKAT (Karoo Array Telescope), the Deep Synoptic Array (DSA), the Australian Square Kilometer Array Pathfinder (ASKAP), the next generation Very Large Array (ngVLA) and the Square Kilometer Array (SKA). Existing facilities such as Arecibo (currently being outfitted with a 40~beam L-band system), Parkes, Effelsberg, the upgraded Westerbork Synthesis Radio Telescope array (WSRT), the Green Bank Telescope (GBT), the VLA, the upgraded Giant Metre Wave Radio Telescope (uGMRT), the Low Frequency Array (LOFAR), and the Long Wavelength Array (LWA) will also be very important.

 
All these facilities will be carrying out pulsar surveys which are in varying states of planning and scope. We provide a brief overview of the expected results\footnote{Yields for any survey can be estimated using the PsrPopPy simulator at {\tt http://psrpop.phys.wvu.edu}.} for some of them based on {\tt PsrPopPy}.  
Survey strategies vary greatly, depending on the 
characteristics of each instrument. 
In Table 1, we summarize the number of normal and MSPs detectable by putative searches with each of the above facilities. The numbers are illustrative in the sense that they are based on the entire sky accessible from each telescope and integration times that may take several years to cover in some cases.
In addition, they do not account for the number of pulsars currently known\footnote{As of March 2019, 2659 radio pulsars are currently listed in the Australia Telescope National Facility (ATNF) catalog \cite{mhth05} which can be accessed at {\tt http://www.atnf.csiro.au/research/pulsar/psrcat}.}. However, they do provide a measure of each telescope's potential as a probe of the Galactic pulsar population and can be scaled and refined accordingly in the future as the experiments are being carried out. As these estimates demonstrate, the radio pulsar population will increase substantially in the coming decade. We anticipate a Galactic sample size approaching $10^4$ pulsars and  $>10^3$ MSPs. Excellent prospects for finding Galactic Center pulsars come from the ngVLA and SKA \cite{2019AAS...23336124D}.

\begin{table}
\centering
\begin{tabular}{lrrrrrr}
Telescope &
Frequency band &
Gain &
Integration &
Sky coverage &
$N_{\rm normal}$ & 
$N_{\rm MSP}$ \\
    &
(GHz) &
(K/Jy) &
(s)   &
($^{\circ}$) &
     &
     \\
\hline
Arecibo  & 1.3--1.8 & 10 & 600  & $0 < \delta < +38$ & 4200 & 850 \\
CHIME & 0.4--0.8 & 2  & 900 & $\delta > -20 $  & 5600  & 650 \\
MeerKAT & 1.1--1.8 & 2.6  & 300  & $-90 < \delta < +40$ & 12000  & 1600  \\

FAST  & 1.1--1.9 & 20 & 20  &  $-20 < \delta < +60$ & 3400 & 1800 \\
ngVLA & 1.2--3.5 & 7.6 & 240 &  $-15<l<265$; $\vert b \vert <1$                    &  2200 &  100   \\
\end{tabular}
\caption{Summary of {\tt PsrPopPy} runs showing the predicted sample sizes of both normal pulsars ($N_{\rm normal}$) and MSPs ($N_{\rm MSP}$) obtained from the online simulator for putative surveys planned for the coming decade. We do not account for overlapping detections, or currently known pulsars. }
\end{table}

\section{Concluding Remarks}

The past 50 years of pulsar astrophysics has informed us that we are only scratching the surface of the neutron star zoo \cite{2018IAUS..337....3K}. Some of the landmark discoveries during this time have included the original double neutron star binary  \cite{ht74}, millisecond pulsars \cite{bkh+82}, pulsar planetary systems \cite{wf92}, the double pulsar \cite{bdp+03,lbk+04}, radio magnetars \cite{crh+06}, fast radio bursts \cite{lbm+07}, ultra-light degenerate companions \cite{2011Sci...333.1717B,2018ApJ...864...15K}, and  triple systems \cite{tat93,tacl99,ajrt96,2014Natur.505..520R,2018Natur.559...73A}.
With such a large yield expected by the upcoming surveys, further rare and exotic discoveries are anticipated. High among these would be the discovery of pulsar--black hole binary systems. These are anticipated in the Galactic field \cite{2018MNRAS.477L.128S}, but also around Sgr~A* \cite{2011MNRAS.415.3951F} and in globular cluster systems \cite{2019arXiv190205963Y}. Such systems could provide highly sensitive test beds of general relativity that go well beyond the already impressive results \cite{ksm+06}. The large numbers of MSPs anticipated will be crucial to the sensitivities of the pulsar timing arrays as they move into an era of low-frequency gravitational wave detection \cite{2014arXiv1409.4579M}.

\end{document}